\documentclass[twocolumn]{aastex7}

\shortauthors{Kotten et al.}
\usepackage{amsmath} 
\usepackage{subcaption}
\usepackage{graphicx}
\usepackage{hyperref}
\usepackage{tablefootnote}

\newcommand\simp{SIMP 0136+0933}

\begin{document}

\title{On the Detectability of Volcanic Exo-Ios That May Fuel Auroras on Super-Jupiters}

\author[orcid=0000-0000-0000-0001]{Brooke Kotten}
\affiliation{Department of Astronomy, University of Michigan, Ann Arbor, MI 48109, USA}
\email{bkotten@umich.edu}  

\author[0000-0002-9521-9798]{Mary Anne Limbach}
\affiliation{Department of Astronomy, University of Michigan, Ann Arbor, MI 48109, USA}
\email{mlimbach@umich.edu}  

\author[0000-0003-0489-1528]{Johanna M. Vos}
\affiliation{ Department of Astrophysics, American Museum of Natural History, Central Park West at 79th Street, NY 10024, USA}
\affiliation{School of Physics, Trinity College Dublin, The University of Dublin, Dublin 2, Ireland}
\email{johanna.Vos@tcd.ie}  

 \author[0009-0008-9086-8095]{Merle Schrader}
 \affiliation{School of Physics, Trinity College Dublin, The University of Dublin, Dublin 2, Ireland}
\email{schradem@tcd.ie}

\author[0000-0003-2015-5029]{Allison McCarthy}
\affiliation{School of Physics, Trinity College Dublin, The University of Dublin, Dublin 2, Ireland}
\email{mccara45@tcd.ie} 

\author[0000-0001-5125-1414]{Melodie M. Kao}
\affiliation{Lowell Observatory, 1400 W Mars Hill Road, Flagstaff, AZ 86001, USA}
\email{mkao@lowell.edu} 

\author[0000-0003-3008-1975]{Mikayla J. Wilson}
\affiliation{Department of Astronomy \& Astrophysics, University of California, Santa Cruz, CA 95064, USA}
\email{mijawils@ucsc.edu} 

\author[0009-0004-7103-7828]{Andrew Householder}
\affiliation{Department of Astronomy, University of Michigan, Ann Arbor, MI 48109, USA}
\email{andhouse@umich.edu}

\author[0000-0001-6098-3924]{Andrew Skemer}
\affiliation{Department of Astronomy \& Astrophysics, University of California, Santa Cruz, CA 95064, USA}
\email{askemer@ucsc.edu} 

\author[0000-0002-7352-7941]{Kevin B. Stevenson}
\email{Kevin.Stevenson@jhuapl.edu}
\affiliation{Johns Hopkins APL, 11100 Johns Hopkins Rd, Laurel, MD 20723, USA}

\author[0000-0002-2011-4924]{Genaro Suárez}
\affiliation{ Department of Astrophysics, American Museum of Natural History, Central Park West at 79th Street, NY 10024, USA}
\email{gsuarez@amnh.org}  

\author[0000-0003-4614-7035]{Beth Biller}
\affiliation{Institute for Astronomy, University of Edinburgh, Royal Observatory, Edinburgh, EH9 3HJ, UK }
\affiliation{Centre for Exoplanet Science, University of Edinburgh, Edinburgh, EH9 3FD, UK}
\email{bbiller@ed.ac.uk}

\author[0000-0001-6251-0573]{Jacqueline Faherty}
\affiliation{ Department of Astrophysics, American Museum of Natural History, Central Park West at 79th Street, NY 10024, USA}
\email{jfaherty@amnh.org}

\author[0000-0003-4636-6676]{Eileen C. Gonzales}
\affiliation{Department of Physics and Astronomy, San Francisco State University, 1600 Holloway Avenue, San Francisco, CA 94132, USA}
\email{egonzales@sfsu.edu}

\author[0000-0002-0638-8822]{Philip Muirhead}
\affiliation{ Department of Astronomy and Institute for Astrophysical Research, Boston University, 725 Commonwealth Ave., Boston, MA 02215, USA}
\email{philipm@bu.edu}

\begin{abstract}
Studies suggest Jupiter's aurorae are supplied with plasma from volcanic outflows on the planet's innermost moon, Io. Repeating bursts of radio emission thought to trace massively scaled-up analogs of Jupiter's aurorae have been detected around nearly a dozen isolated substellar worlds, yet the source of the electrons fueling the aurorae remains unknown. Volcanism from tidally heated exosatellites may provide the plasma that fuel the aurora on these worlds.
We assess whether transit observations provide a viable means of detecting exosatellites around aurorally active substellar worlds, thereby enabling future tests of this hypothesis.
Specifically, we analyze JWST near- and mid-infrared light curves of \simp{}, a $12.7 M_J$ ``super-Jupiter", known to exhibit auroral emission. 
We demonstrate the capability to detect exosatellites in the \simp{} system with satellite-to-host mass ratios comparable to those of Jupiter's Galilean moons, achieving detection success rates of 66\% for Io-to-Jupiter mass ratio satellites and 93\% for Ganymede-to-Jupiter mass ratio satellites. 
Although the existing light curve is sufficient to demonstrate that this technique is capable of detecting transiting exosatellites, the available archival data are too short in duration to place meaningful constraints on the presence of a transiting satellite in this system. We conclude that JWST light curves spanning $\sim$1.5~days for 4-12 known aurorally active super-Jupiters would be sufficient to yield evidence for or against this hypothesis. A small target sample may suffice, as short satellite periods boost transit probabilities and aurorally active worlds may be preferentially observed near edge-on inclinations.

\end{abstract}

\keywords{Transits, Natural satellites (Extrasolar), Aurorae, Volcanism, Free floating planets, Brown dwarfs, Magnetospheric radio emissions, Io}

\section{Introduction}
\label{sec:introduction}

Repeating radio bursts caused by massively scaled-up analogs of Jupiter’s aurorae have been detected around a growing sample of brown dwarfs and planetary mass objects \citep{2005ApJ...627..960B, 2007ApJ...663L..25H, 2008ApJ...684..644H,2012ApJ...747L..22R, Melodie_2016_auroraLTdw,2017ApJ...834..117W,Vedantham_2023_tdwarfaurora,Rose_2023_Tdwarfradio,Melodie_2024_radiooccurence,Guiradio_2025}.
The mechanisms driving substellar aurorae remain unconfirmed \citep{Hallinan2015, pineda_2017}, and for these isolated substellar objects that lack a host star, only mechanisms that do not require a stellar wind can plausibly occur \citep{Melodie_2016_auroraLTdw, pineda_2017, Kao_2018_0136aurora, saur_2021}.
One suggested possibility is that they are analogous to mechanisms occurring on Jupiter \citep{nichols2012, pineda_2017, turnpenney2017, saur_2021}. 

Orbiting moons can perturb Jupiter’s magnetic field to excite aurora \citep[i.e., the Jupiter-Io current system;][]{1996Sci...274..404C,2001AdSpR..27.1915B, 2001P&SS...49.1067C, 2018Sci...361..774M}. 
Plasma ejected from Io is guided along Jupiter's magnetic field lines into the planet's polar regions, where it precipitates into the upper atmosphere, producing aurorae and radio emission. 
Although the observed substellar aurorae are $>1000\times$ more powerful than those seen on Jupiter \citep{Melodie_2016_auroraLTdw, Vedantham2020,2023Natur.619..272K, 2023Sci...381.1120C}, modeling suggests that the same mechanism responsible for Jupiter's aurorae could produce the flux densities of aurorae observed on substellar worlds \citep{nichols2012}. However, this would necessitate a sustained and substantial source of plasma. 
In the case of Jupiter, the moon Io supplies $\sim$98\% of the plasma in the Jovian magnetosphere \citep{hill1983}.  %see table 10.1 

Vigorous volcanic activity on an orbiting Io-analog is a plausible source to supply the electrons driving aurora on isolated substellar worlds \citep{Hallinan2015, pineda_2017, faherty2024Natur.628..511F}. 
Furthermore, close-in exosatellites (i.e., small companions to substellar objects) may plausibly occur at high rates around substellar hosts, given the elevated occurrence rates observed just above and below the substellar regime; specifically, at the low-mass-host end, theoretical formation models reproduce the high occurrence rate of moons in the Solar System \citep{canup2006, cilibrasi2021, limbach_2023_tempo1}. At the high-mass end, observations of mid- to late-M dwarfs \citep{hardegreeullman2019} support the hypothesis that multiple close-in exosatellites may be common around substellar and planetary-mass hosts.

Until now, we have lacked observatories with the sensitivity required to search for exo-Ios that may undergo vigorous tidal heating around aurora-emitting substellar objects.
Previous surveys for companions to substellar worlds, including transit searches (\citealt{matthias_2016, Limbach_2021, Limbach_2024_occurance}; PINES \citealt{tamburo_muirhead_2019_pines, tamburo2022I, tamburo2022II}), doppler monitoring \citep{vanderburg_2018,ruffio_2023} and direct imaging \citep{lazzoni_2020} have generally been sensitive only to satellites larger than $\sim$sub-Neptunes \citep[e.g.,][]{2026A&A...705A.217K} and have lacked the precision necessary to detect small terrestrial exosatellites analogous to Jupiter’s Io.

However, recent observations with JWST have demonstrated for the first time that exosatellites in the Io-analog regime are now within reach, via searches for exosatellite transits in the light curves of substellar worlds \citep{Householder2025, wilson2025}.
One drawback of the transit method is that it can only detect satellites in systems that are viewed close to edge-on. 
Fortunately, brown dwarfs and planetary mass objects with strong detected aurora \citep[e.g.][]{Curiel_2020_radioplanet, Guiradio_2025, hallinan2008ApJ...684..644H} are more likely to be detected at intermediate to equatorial viewing angles, as their radio emission is beamed and therefore disfavored in near pole-on geometries \citep{pineda_2017}. This is consistent with observations of Jupiter, whose radio emission is preferentially detected near the equatorial plane \citep{carr1983phjm.book..226C, Ladreiter_Leblac_1989}. 
Additionally, such satellites must reside on short-period orbits to sustain sufficient tidal heating and plasma production, which further increases their transit probabilities \citep{Winn_2010} and reduces the light curve duration required for transit detection. 

In this paper, we assess if the transit technique is capable of detecting exo-Ios orbiting aurorally active substellar worlds. 
We demonstrate the technique with JWST light curve data for SIMP J013656.5+093347.3 (hereafter \simp), a young ($200\!\pm\!50$~Myr old; \citealt{gagne_2017_simp0136discovery}), free-floating planetary mass ($12.7~M_J$) object. 
It is radio loud, a property attributed to auroral activity \citep{Melodie_2016_auroraLTdw, Kao_2018_0136aurora}, and has six hours of JWST/MIRI and NIRSpec spectroscopic time series observations that were originally obtained for variability studies and presented in \citet{mccarthy_2025}, \citet{nasedkin_2025} and M. Schrader et al. (2026, submitted). %The JWST weather report: Unravelling the atmospheric variability of isolated worlds using Principal Component Analysis
Identifying information and properties of \simp{} can be found in Table \ref{tab:host_params}. 

We outline our paper as follows. Section \ref{sec:data/obs} details the observational data of \simp. 
In Section \ref{sec:analysis}, we build the light curves, assess the probability of a transit, and inject and recover transits in the light curves. 
In Section \ref{sec:results} we assess if Io-analog transits are detectable in the light curves from the injection-recovery testing.
Section \ref{sec:discussion} evaluates the performance of JWST MIRI and NIRSpec for our satellite transit search technique and describes the ideal survey capable of testing whether exo-Ios fuel aurora on substellar worlds.
Section \ref{sec:conclusions} contains our conclusions.

\section{Data/Observations}
\label{sec:data/obs}

As a part of the JWST Cycle 2 program GO 3548 (PI: Vos), \simp{} was observed for a total of 6 hours, split evenly between Near Infrared Spectrograph (NIRSpec) and the Mid-Infrared Instrument (MIRI) Low Resolution Spectrograph (LRS). 
These observations spanned two consecutive full rotation periods of \simp{} (P=2.4\,hrs); first with NIRSpec/PRISM 0.6--5.3~$\mu$m, R$\sim$100 for 2.88\,hrs and followed by MIRI/LRS 5--14~$\mu$m, R$\sim$100 for 2.98\,hrs. 
NIRSpec Bright Object Time Series observations and MIRI/LRS Time Series Observations were carried out on Jul 23 and 24 of 2023. The JWST Exposure Time Calculator (ETC) predicted SNRs of approximately $1000$--$5000\,\mathrm{hr}^{-1}$ for NIRSpec and $200$--$1000\,\mathrm{hr}^{-1}$ for LRS at full spectral resolution (although the SNR decreases toward longer wavelengths for LRS), which is consistent with the achieved full-resolution performance reported by \citet{mccarthy_2025}.

%I USED THESE FILES: 
% nirspec_file = "SIMP0136LCs/Merle_Schrader/jw03548003001_03101_00001-seg001_nrs1_x1dints_fromIndiv.fits"  
% miri_file = "SIMP0136LCs/MIRI-LRS-data/miri_lrs_tso_stage3_x1dints_toshare.fits"

The MIRI/LRS data were reduced using the JWST STScI pipeline version 1.14.0 Calibration Reference Data System (CRDS) version 11.17.19 and CRDS context file \texttt{jwst 1253.pmap}. As described in \citet{mccarthy_2025}, the JWST Pipeline Stage 1 was run with default settings, while Stage 2 was run with an added custom background subtraction. Full details can be found in \citet{mccarthy_2025}.

The NIRSpec/Prism data were reduced using a modified JWST Stage~2 reduction described in \citet{nasedkin_2025}. Each integration was processed individually rather than applying static calibrations across the full time-series dataset. This approach allowed the \texttt{nsclean} correction for $1/f$ noise to be applied and enabled the derivation of a separate spectral trace for each integration to capture time-dependent shifts from pointing variations. Flat field and photometric calibrations were also applied on a per-integration basis.

\begin{deluxetable}{lcc}
\tablecaption{Properties of SIMP J013656.5+093347\label{tab:host_params}}
\tablehead{
\colhead{Parameter} & 
\colhead{Value} & 
\colhead{Ref}}
\startdata
R.A. & 01:36:56.62 & 1 \\
Decl. & +09:33:47.3 & 1 \\
$\mu_\alpha \cos \delta$ (mas yr$^{-1}$) & $1222.70\pm0.78$ & 1 \\
$\mu_\delta$ (mas yr$^{-1}$) & $0.5 \pm 1.2$ & 1 \\
Distance (pc) & $6.139\pm0.037$ & 1 \\
Radius ($R_{\rm Jup}$) & $1.22\pm0.01$ & 2 \\
Age (Myr) & $200\pm50$ & 2 \\
Mass ($M_{\rm Jup}$) & $12.7 \pm 1.0$ & 2 \\
${\rm T_{\rm eff}}$ (K) & $1098\pm6$ & 2 \\
$\log g$ (dex) & $4.31\pm0.03$ & 2 \\
Rotation Period (hr) & $2.414 \pm 0.078$ & 3 \\
Inclination ($^\circ$) & $80^{+10}_{-12}$ & 4 \\
Spectral Type & T2.5$\pm$0.5 & 5 \\
$K$ (UKIDSS, mag) & $12.585\pm0.002$ & 6 \\
$J$ (2MASS, mag) & $13.46\pm0.03$ & 7 \\
Moving Group Member & Car-N & 2 \\
\enddata

\tablerefs{
(1) \cite{2017ApJ...850...11D};
(2) \cite{gagne_2017_simp0136discovery};
(3) \cite{yang_2016ApJ};
(4) \cite{vos_2017_viewinggeoBDs};
(5) \cite{artigau2006};
(6) \cite{2007MNRAS.379.1599L};
(7) \cite{2mass_2006AJ....131.1163S}.
}

\end{deluxetable}

\section{Analysis}
\label{sec:analysis}

\subsection{Creating Light Curves}
\label{subsec:lcs}
Our goal is to use the \simp{} observations to construct light curves and perform transit injection–recovery tests to assess whether exosatellites analogous to Io would be detectable around aurorally active substellar worlds. To begin this process, we first construct the light curves used in our transit search.

\begin{figure*}
    \centering
    \includegraphics[width=\textwidth]{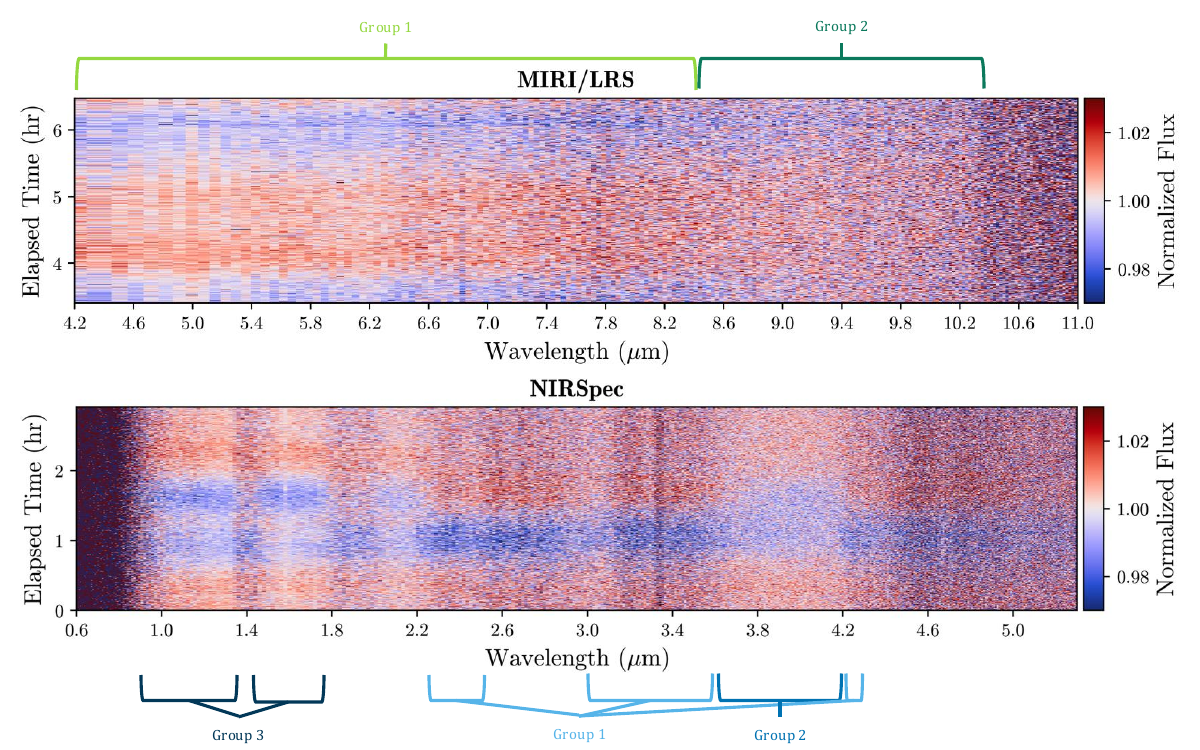}
    \caption{Unbinned spectrophotometric MIRI (top) and NIRSpec (bottom) data for \simp{}. Multiple variability patterns are distinguishable by eye, particularly in the NIRSpec data. The wavelength grouping is used to build the light curves shown in Figure~\ref{fig:bothlcs} (grouping is approximate and displayed for visual purposes only).}
    \vspace{-0mm}
    \label{fig:bothvars}
\end{figure*}

Variability maps for MIRI and NIRSpec display the normalized flux of \simp{} as a function of wavelength and time in Figure \ref{fig:bothvars} for our time series observations. 
While we observe variability across all wavelengths, certain wavelength regions may exhibit similar behaviors that correspond to different atmospheric pressures and physical processes on the substellar world, resulting in distinct variability patterns \citep[see][]{biller_2024, chen2025MNRAS.539.3758C,mccarthy_2025, oliverosgomez_2026ApJ...997..136O}. 

One challenge we encounter when searching for satellites around substellar worlds is that intrinsic host variability may mimic a transit signal \citep{Limbach_2021}. However, substellar variability is typically chromatic (ex. \citealt{biller_2024, mccarthy_2025}), whereas transits are (to first order) achromatic \citep{Householder2025}. Leveraging this differing behavior, we create multiple light curves from the data by sorting into wavelength groups with similar variability features. This will enable us to simultaneously search multiple wavelength-grouped light curves, more effectively distinguishing transit signals from intrinsic variability.
Including multiple light curves with different variability patterns reduces the chance of mistaking variability for a transit signal, enabling the detection of small satellites \citep{Householder2025, wilson2025}. 

\setcounter{footnote}{0}
There are several possible methods for constructing spectrally binned light curves with similar variability behavior. \citet{mccarthy_2025} applied a $k$-means clustering algorithm to group wavelength ranges of \simp{} MIRI and NIRSpec data with similar temporal behavior into light curves. For our transit search we select only a few groups influenced by the \citet{mccarthy_2025} clustering.\footnote{In this work \emph{clustering} refers to selections made using a clustering algorithm (e.g. \citealt{mccarthy_2025}) whereas \emph{grouping} denotes a manual process of identifying patterns to select orthogonal light curves for the transit search, although grouping may be informed by clustering results.} We tested the sensitivity of our results to the chosen wavelength groupings by varying the bin boundaries by $\sim$10\%. This produced no measurable change in our ability to detect transits in the light curves. We therefore conclude that grouping ``by eye,'' rather than employing a formal clustering identification algorithm, is sufficient for our purposes.

\begin{figure*}
    \centering
    \includegraphics[width=0.95\textwidth]{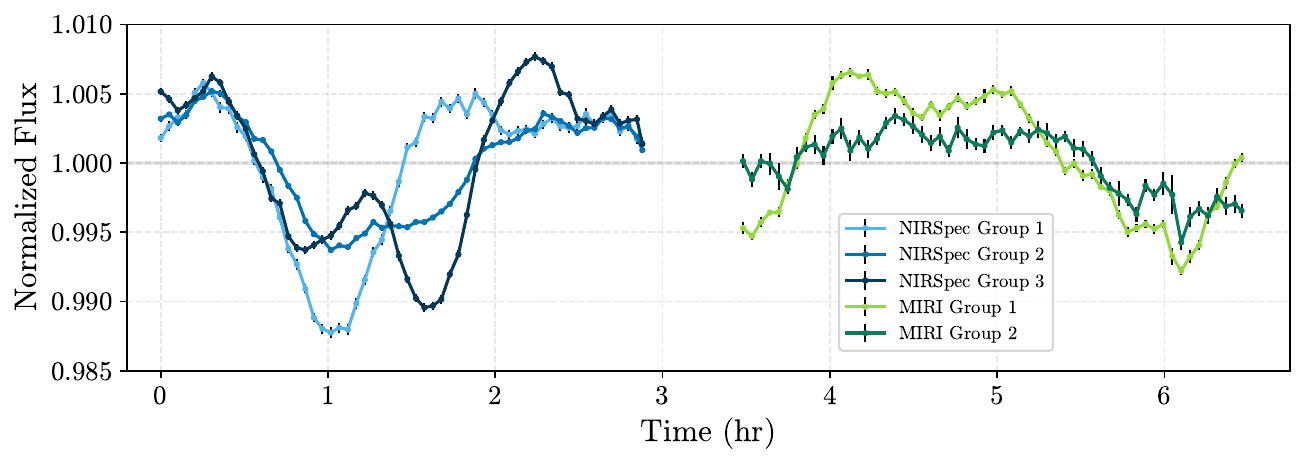}
    \caption{\simp{} MIRI and NIRSpec light curves constructed for the exosatellite transit injection-recovery testing conducted in this study. Note that the larger amplitudes and more complex variability structure at shorter wavelengths are typical of substellar atmospheres \citep[e.g.][]{mccarthy_2025}. The grouping corresponds to Figure \ref{fig:bothvars} and Section \ref{subsec:lcs}. As pointed out by \citet{mccarthy_2025}, two full rotation periods can be viewed from 1--3.4\,hrs and 3.4--5.8\,hrs.}%COMMENT: not adding wavelength ranges to legend because NIRSpec ranges are not singular
    \vspace{-0mm}
    \label{fig:bothlcs}
\end{figure*}

Figure \ref{fig:bothlcs} provides light curves of the selected MIRI groups in green. There are two distinct variability patterns visible in the MIRI data, and our ``by-eye'' grouping is similar to two algorithm-identified clusters found by \citet{mccarthy_2025}. The first wavelength group spans $4.2-8.5~\mu m$. The second wavelength group spans $8.5-10.3~\mu m$ and probes the top of the forsterite cloud layer. 
Beyond 10.3~$\mu m$, the SNR was too low to be a useful inclusion to a light curve. 

While the nine NIRSpec clusters identified by \citet{mccarthy_2025} span a wide range of variability behaviors, including all clusters in our simultaneous transit search is computationally prohibitive, as runtime scales by a factor of $\sim 10$ per additional light curve. We therefore constructed three representative light curves that capture the dominant variability modes.
Figure \ref{fig:bothvars} illustrates our NIRSpec grouping choices in blue and the final three light curve groups are shown in Figure \ref{fig:bothlcs}.
The wavelength group 1 light curve utilizes \citet{mccarthy_2025}'s clusters 1 and 2, corresponding to 2.2515-2.4807, 3.0790-3.0919, 3.1177-3.1432, 3.1558-3.5154, 3.5267-3.5830, and 4.2172-4.3289$~\mu m$. 
The group 2 light curve is cluster 6 from \citet{mccarthy_2025}, spanning 3.6384-4.2078$~\mu m$; it is mostly flat and probes the forsterite cloud layer. 
The group 3 light curve combines \citet{mccarthy_2025}'s clusters 8 and 9, corresponding to 0.9050-1.3188, 1.3382-1.3577, and 1.4567-1.7364$~\mu m$; it probes the highest pressure levels near the iron cloud layer and displays two troughs. 

The light curves are constructed as follows: 1) the flux in each wavelength group is normalized by its mean value; 2) for each wavelength group, the light curve is computed as the median flux across all channels in the group; and 3) the resulting wavelength grouped light curve is temporally binned to the desired cadence of $\sim$3~min using the mean flux within each time bin. The resulting light curves are binned to 58 points for NIRSpec (reduced by a factor of 100) and 57 points for MIRI (reduced by a factor of 10). This temporal binning substantially reduces the computational cost of the injection-recovery process and does not impede transit detectability, as the native cadence of the observations is much shorter than the minimum expected transit duration (1.83 seconds for NIRSpec and 19.25 seconds for MIRI). Furthermore, this binning procedure allows us an independent estimate of the flux uncertainty.

The uncertainty associated with each binned point is calculated using the standard error of the mean flux values across all wavelengths within the group ($\sigma/\sqrt{\text{n}}$ where $n$ is the number of native time samples combined into each binned point (10 points for MIRI and 100 for NIRSpec). This results in nearly identical temporal sampling of $\sim$3~min, since the native NIRSpec cadence is higher than that of MIRI. The associated uncertainties are shown in Figure \ref{fig:bothlcs} and are 213--359ppm/3min for NIRSpec and 323--553ppm/3min for MIRI.

\subsection{Expected Transit Properties and Probability}
\label{subsec:parameterspace}

We now establish the parameter space in which we should search for transiting exosatellites, beginning with the range of physical and orbital parameters expected for exo-Ios. These properties will inform our transit search algorithm’s priors. 

\subsubsection{Transit Duration}
\label{subsubsec:duration}

Following \cite{Winn_2010}, the transit duration is given by
\begin{equation}
\label{tdurEQ}
    t_{\rm dur} =\frac{P}{\pi}\sin^{-1}{\left[\frac{R_h}{a}\frac{\sqrt{(1+k)^2-b^2}}{\sin{i}}\right]},
\end{equation}
where $P$ is orbital period, $a$ is semi-major axis, $k=\frac{R_{sat}}{R_h}$, $R_h$ is host radius, $R_{sat}$ is satellite radius, $i$ is inclination, and $b$ is the impact parameter. 
The smallest physically allowed orbital separation is set to the Roche limit:
\begin{align*}
    d_R = 2.44*R_h\left( \frac{\rho_h}{\rho_{sat}}\right)^{1/3},
\end{align*} %no need to number
where $\rho_h$ is host density and $\rho_{sat}$ is exosatellite density. Throughout this paper, we will use mass-radius relations from \citet{chen_kipping_2017}, which result in $\rho_{sat} \approx \rho_\oplus$ for this regime. 
The Roche limit for \simp{} is $d_R = 0.00166\,\mathrm{au}$, or 2.85 $R_{h}$. 
%which is roughly 1.5 times the distance from the Earth to its moon. Though, the Moon orbits at a distance 20 times the Earth's Roche limit

Using Equation~\ref{tdurEQ}, we compute the possible transit durations for an exosatellite orbiting \simp{} at orbital separations spanning $a = d_R$ to $a = 0.2\,\mathrm{au}$ ($\sim$3--350\,$R_{host}$). %precise numbers are 2.84-343
We assume that satellites at separations $a > 0.2\,\mathrm{au}$ lie well beyond the regime in which significant tidal heating would be expected. For comparison, Io orbits at $a = 0.003\,\mathrm{au}$ ($6\,R_{J}$), and tidal heating scales as $a^{-15/2}$ \citep{2013ApJ...769...98P}.

\begin{figure}
    \centering
    \includegraphics[width=0.9\linewidth]{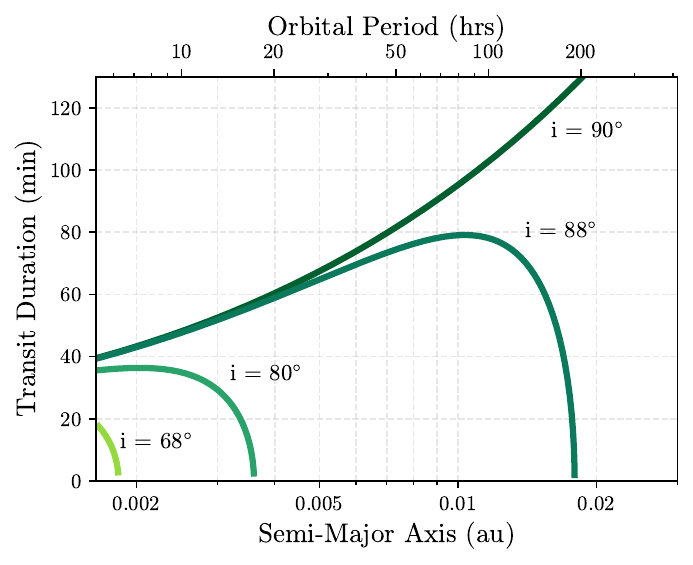}
    \caption{Transit durations as a function of semi-major axis for satellites orbiting \simp{}, assuming orbital inclinations ranging from $i = 68^\circ$ to $90^\circ$. Typical transit durations span $\sim$10-100 minutes.} %I cannot add Io and others for reference because of the orbital period and semi major axis calculations
    \label{fig:durations}
\end{figure}

The resulting transit durations are calculated for varying viewing inclinations and plotted in Figure \ref{fig:durations}. 
The measured viewing inclination of \simp{} is constrained to $i = 80^{+10}_{-12}\,^\circ$ \citep{vos_2017_viewinggeoBDs}, or nearly equator-on. Assuming spin–orbit alignment between the exosatellite orbit and the substellar object spin axis, we plot the transit durations over a range of inclinations within $1\sigma$ of the measured spin-axis inclination of \simp{}.
Transit durations are highly dependent upon the viewing inclination of the satellite system, where near edge-on/equatorial orientations have the longest transits.
For aurorally active worlds, the key geometric constraint is that they are unlikely to be viewed pole-on, as the beaming effect would render their emission difficult to detect in such configurations. Instead, geometric selection effects favor viewing inclinations closer to equator-on ($i \approx 90^\circ$), though not necessarily exactly equator-on, as discussed in Section~\ref{sec:introduction}.
Based on this analysis, we adopt a prior of $8\,\mathrm{min}<t_{duration}<120\,\mathrm{min}$ for our transit search.

\subsubsection{Transit Depth}
\label{subsubsec:depth}
The transit depth depends upon the radius of the substellar host and satellite, and is given by 
\begin{equation}
    \delta_{transit} = \left(\frac{R_{sat}}{R_{h}}\right)^2.
\end{equation}
For \simp{} ($R_h = 1.22 R_J$, \citealt{gagne_2017_simp0136discovery}) we may expect transit depths of $\sim$5350\,ppm (0.54\%) for an Earth-sized object, $\sim$2800\,ppm for a Ganymede-sized object, $\sim$1500\,ppm for a Mars-sized object, $\sim$860\,ppm for a Titan-sized object, and $\sim$450\,ppm for an Io-sized object. 

However, host to satellite mass ratios may be a better comparison metric than radius relations. Observations of solar system satellites and theoretical simulations find that satellite masses are generally expected to scale linearly with host mass \citep{canup2006}. More explicitly, satellites orbiting hosts more massive than Jupiter are likely to be correspondingly more massive, while radius relations of objects with $M_h>1M_J$ will remain the same. Consequently, when considering the detectability of Io analogs around super-Jovian hosts, it is reasonable to require sensitivity to satellites with satellite-to-host mass ratios comparable to those of Jupiter’s Galilean moons ($\sim$\,a few times $10^{-5}$), rather than to satellites with identical absolute masses \citep{2016Icar..266....1M,2018MNRAS.475.1347M,rufu2025NatCo..16.4853R}. 
A Jupiter-Io mass analog ($\frac{M_{Io}}{M_{Jup}}=4.7\times10^{-5}$) satellite orbiting \simp{} would have a transit depth of 1770\,ppm and be equivalent to a Mars-sized object orbiting \simp{}. In conversions between mass and radius throughout this work, we use a terrestrial mass-radius relation of $R\sim M^{0.28}$ from \citet{chen_kipping_2017}, which results in near-Earth densities for exosatellites. 

We constrain our search to satellites $<1.1\,\text{R}_\oplus$ (6500\,ppm, 1.4\,$\text{M}_\oplus$) because exosatellite formation theory \citep{canup2006,inderbitzi2020MNRAS.499.1023I, liu2020A&A...638A..88L, cilibrasi2021, limbach_2023_tempo1, rufu2025NatCo..16.4853R} and observational evidence from mid- to late-M-dwarfs \citep{hardegreeullman2019} predicts that companions larger than this will be rare in the super-Jovian to low brown dwarf mass regime. 

\subsubsection{Geometric Transit Probability}
\label{subsubsec:geo_prob}
In addition to the parameter space that we expect for transits, our likelihood of actually detecting a transit in the data depends on the geometrical probability that a transit will occur from our viewing angle. For a substellar object and satellite system (the same as a star-planet system) at a random inclination, the geometric transit probability is $R_h/a$ (when $R_h\gg R_{sat}$ and $e=0$) \citep{Winn_2010}. 

Assuming that the substellar host has a known inclination, specifically between a lower limit inclination $i_l$ (as radio-loud substellar worlds are not view near pole-on) and 90$^\circ$ and further assuming spin-orbit alignment between the exosatellite orbit and the host’s spin axis \citep{2010ApJ...712.1433B,Limbach_2021}, 
we can compute the transit probability for an exosatellite. If it is known to be between an inclination of $i_l$ to 90$^\circ$, we derive that the geometric transit probability would be 
\begin{equation} \label{eqn:Pgeo}
P_{\mathrm{tra,\,geo}} =
\min\!\left[
1,\;
\frac{R_h}{a}\,
\frac{1}{\sin\!\left(\dfrac{90^\circ - i_l}{2}\right)}
\right],
\end{equation}
where the $\min[\cdot,\cdot]$ function returns the smaller of its two arguments, so the transit probability is taken as the geometric term unless it exceeds unity, in which case it is capped at 1 to ensure it remains a physical probability. 
While a system observed at exactly $i=90^\circ$ would have a 100\% transit probability for any given $a$, even a slight deviation such as $i=89^\circ$ introduces a dependence on $a$. 

\begin{figure}
    \centering
    \includegraphics[width=0.9\linewidth]{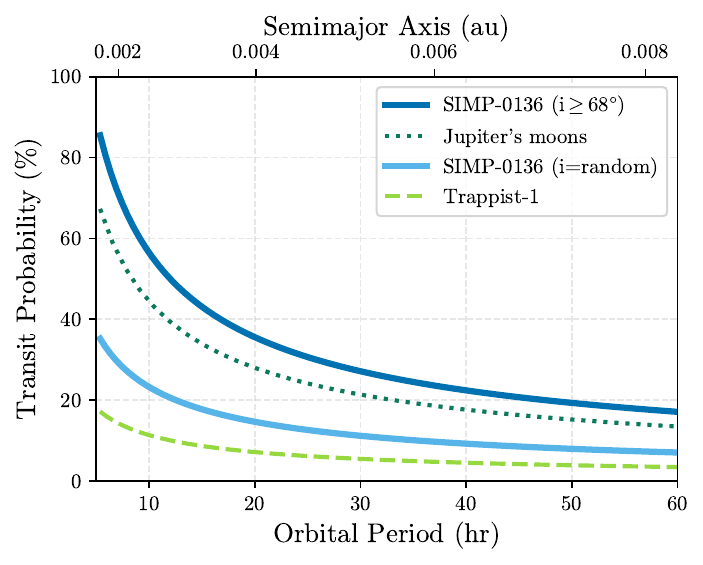}
    \caption{Geometric transit probability ($P_{tra, \,geo}$, Equation\,\ref{eqn:Pgeo}) for \simp{}, Jupiter and TRAPPIST-1, assuming random system inclinations, and \simp{} with a measured inclination (assuming spin-orbit alignment). Over this host-mass range, transit probabilities are generally higher for lower-mass hosts. When the inclination constraint for \simp{} is included, the probabilities increase further, reaching 25--60\% for orbital periods of 10--30~hr, indicating that a substantial fraction of such systems would permit viewable transits of short-period satellites. The reference semi-major axis values correspond to the host mass of \simp{}. }
    \label{fig:geo_prob}
\end{figure}

Figure \ref{fig:geo_prob} displays the geometric transit probabilities for several systems (assuming random viewing inclinations; i.e., $P_{tra, \,geo} = R_h/a$) and \simp{} using Equation~\ref{eqn:Pgeo} given its measured lower inclination limit of $i_l = 68^\circ$ (Table \ref{tab:host_params}) and assuming spin-orbit alignment. For random viewing angles in this substellar and ultracool dwarf host mass range, exosatellite transit probabilities scale inversely with host mass: Jupiter-host-mass systems have the highest transit probabilities, \simp{} lies in the intermediate regime, and TRAPPIST-1 has the lowest \citep[see Figure 3 in ][]{Limbach_2021}. This implies that, if each system hosted a single companion at a given orbital separation $a$, substantially fewer Jovian-like systems than TRAPPIST-1 systems would need to be monitored to identify one in which the companion transits, simply because the transit probability is much higher for the Jovian-like case.

As shown in Figure \ref{fig:geo_prob}, when the inclination constraint for \simp{} is included, the transit probability becomes even higher than that of a randomly oriented Jovian system. For example, if every system hosted an exosatellite with a 20-hour orbital period, the resulting transit probability for \simp{} is $\sim$36\%, meaning that, statistically, observing only three \simp{}-like systems for 20 hours each should be sufficient to expect $\sim$one transiting exosatellite.
We also note that, if a system is observed for a duration shorter than the satellite’s orbital period, the probability of detection is further reduced (see Section~\ref{subsec:results_probs}).

\subsection{Transit Injection-Recovery Methodology}
\label{subsec:injrec_methods}

To explore if transiting Io-analogs would be detectable in the light curves of the aurorally active \simp{}, we conduct transit injection and recovery testing using the light curves built in Section \ref{subsec:lcs}.

\subsubsection{Injections}
\label{subsubsection:inj}
To determine our sensitivity to transiting exosatellites, we inject transits into the light curves that are within the following parameter space:
\begin{itemize}
    \item Transit mid-times (defined as the mid point in time between transit ingress and egress) between 15 minutes and 2.4 hours ($\sim$\,30\,min before the end of the available light curve). %2.87891402 NIRSPec 2.98049048 MIRI%
    \item Orbital separation between the Roche limit 0.001656\,au and 0.02\,au randomly distributed in log space. The outer limit was chosen to reflect the expected transits for our target (see Figure \ref{fig:durations}).
    \item Uniformly random inclinations within the range of $i=69-90^\circ$, where an inclination of 69.4$^\circ$ is the lowest inclination that would geometrically transit. Coincidentally, this matches the measured spin-axis of \simp{}, though this did not inform our injection parameters. It simply marks the inclination beyond which transits outside the Roche limit no longer occur, so extending the injection range would not yield observable transits.  
   \item We began injecting transit depths of 6500\,ppm and decreased the transit depth until the recovery rate flattened to a few percent. This results transit depths from 100--6500\,ppm ($0.14-1.10\,\text{R}_\oplus$).
\end{itemize}

The above parameters fully describe a trapezoidal transit model but do not include higher-order effects such as limb darkening. Previous studies have shown that the inclusion of limb-darkening parameters does not significantly affect transit detectability within the parameter space explored here \citep{Householder2025}.
We inject 100 transits at 13 different transit depths in each mode, resulting in 2600 total injected transits between MIRI and NIRSpec.

\subsubsection{Recoveries}
\label{subsubsection:recovery}

Our transit detection algorithm uses a Gaussian Process (GP) plus a transit model, fitting the variability and transit signal simultaneously in each light curve. For each instrument (MIRI and NIRSpec independently), all associated light curves are fit together in a single joint fit.
As our light curves are so short in duration, it is not physically plausible that there could be more than one transit during the observation. During this simultaneous fit, we require that the transit fit in each of the light curves is the same, but allow a different GP model to capture the variability's chromaticity. This approach fits chromatic variability across light curve groups simultaneously (with the GP) and still enables achromatic transits to be identified. 

Our detection algorithm fits four transit parameters to all the light curves simultaneously: transit mid-time, transit depth, transit duration, and transit ingress/egress slope. We use the slope of the ingress/egress as a proxy for impact parameter which reduces computational cost. The transit prior ranges used in our search algorithm are shown in Table \ref{tab:searchparams}. The ranges are chosen to reflect the parameter space discussed in Section \ref{subsec:parameterspace}. 
The GP quasi-period kernel based on \citet{vaneylen2018MNRAS.478.4866V} individually fits each light curve with six parameters to account for variability and is described in greater detail in \citet{Limbach_2021} and \citet{Limbach_2024_occurance}. 
Thus, the two MIRI light curve groups have 16 free parameters and the three NIRSpec light curve groups have 22 free parameters.

To sample light curve fits, we use \texttt{dynesty.NestedSampler} \citep{speagle2020_dynesty} with \texttt{nlive = 500} for both NIRSpec and MIRI. 
Although \texttt{Dynesty} recommends adding 50 live points for each parameter\footnote{\url{https://dynesty.readthedocs.io/en/v2.1.5/quickstart.html\#live-points} }, increasing the number of live points beyond 500 did not significantly improve our performance in either case (based on rerunning the same injection recoveries). 
Additionally, the number of live points scales linearly with the algorithm's runtime. Thus, \texttt{nlive=500} optimizes the computational time burden while allowing the highest threshold of sensitivity for our object. We run \texttt{Dynesty} until a precision of \texttt{dlogz = 1} in the Bayesian evidence, where \texttt{z} is the likelihood function.

\begin{table}[t]
\centering
\caption{Transit Search Prior Ranges}
\label{tab:searchparams}
\begin{tabular}{lcc}
\hline
Parameter & MIRI & NIRSpec \\
\hline
Transit depth (ppm) & $\mathcal{U}$(0,7000) & $\mathcal{U}$(0,7000) \\
Transit duration (min) & $\mathcal{U}$(8,120) & $\mathcal{U}$(8,120) \\
Ingress/Egress slope & $\mathcal{U}$(0.001,1.000) & $\mathcal{U}$(0.001,1.000) \\
Transit mid-time (hr) & $\mathcal{U}$(0,2.48)$^*$ & $\mathcal{U}$(0,2.88) \\
\hline
\end{tabular}
% \footnotetext{$^*$For MIRI, the final 30 minutes of data are excluded to avoid the step discussed in Section~\ref{subsec:theblip}.}
\raggedright
{\footnotesize $^{*}$ For MIRI, the final 30 minutes of data are excluded to avoid the step discussed in Section~\ref{subsec:theblip}.}
\end{table}

\begin{figure}
    \centering
    \includegraphics[width=0.95\linewidth]{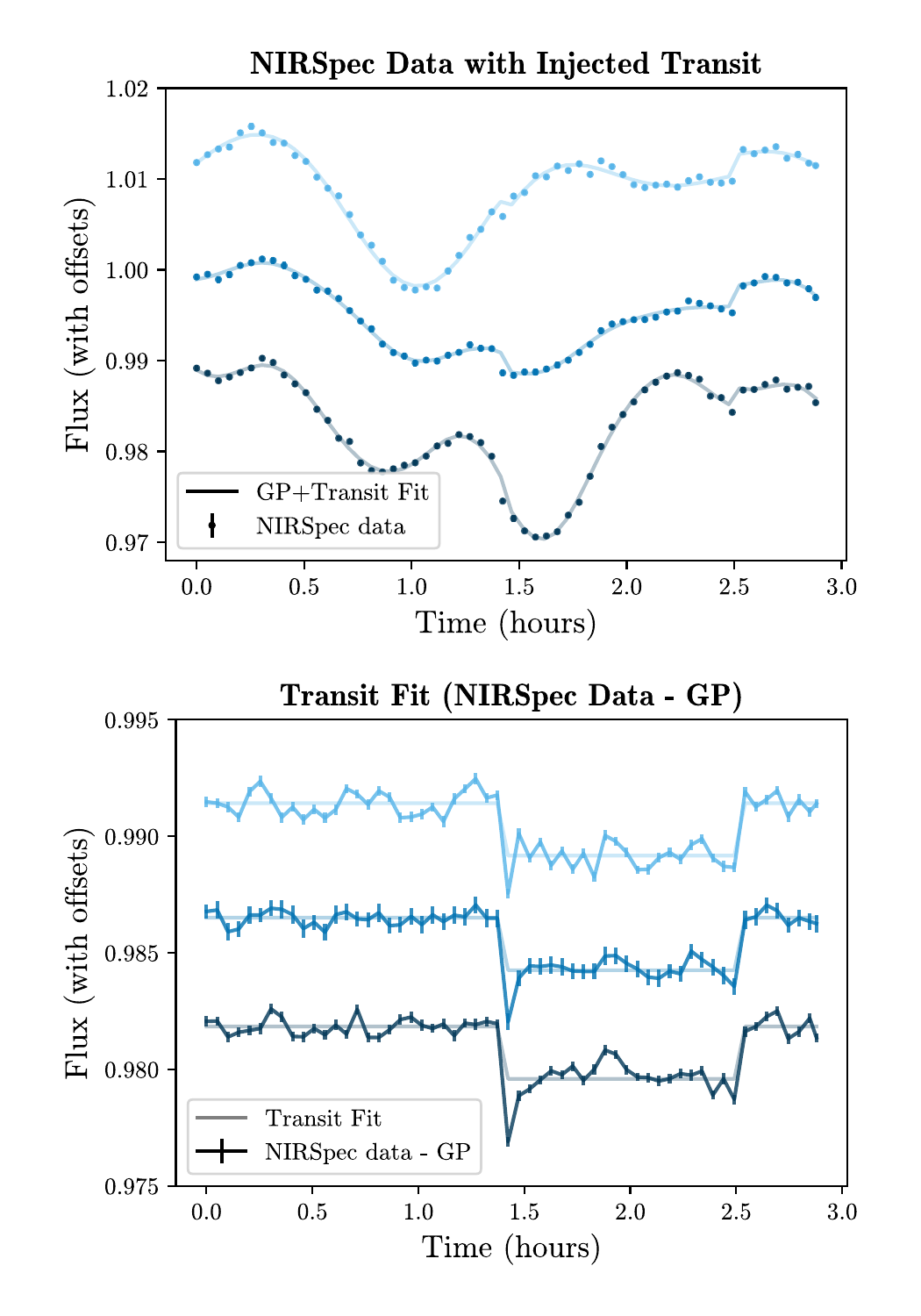}
    \caption{NIRSpec light curves corresponding to grouping in Figure \ref{fig:bothlcs} with a 3000\,ppm ($0.75\,\text{R}_\oplus$) injected transit. The top panel displays the temporally binned light curve with GP fit plus the injected transit and its fit. The bottom panel subtracts the GP fit and displays the transit fit. Offsets between the two panels are unrelated chosen to reflect data with the most visual detail.}
    \label{fig:injected_transit_lc}
\end{figure}

Figure \ref{fig:injected_transit_lc} displays an example of an injected and subsequently recovered transit. The top panel shows the binned NIRSpec light curve groups with an injected 3000\,ppm ($0.75\,\text{R}_\oplus$) transit; while the variability is strong, the large injected transit is visible by eye when looking across the groups. The bottom panel removes the variability fit by the GP and displays the transit fit. 

\begin{figure*}[ht]
    \centering
    \begin{subfigure}[b]{0.329\linewidth}
        \centering
        \includegraphics[width=\linewidth]{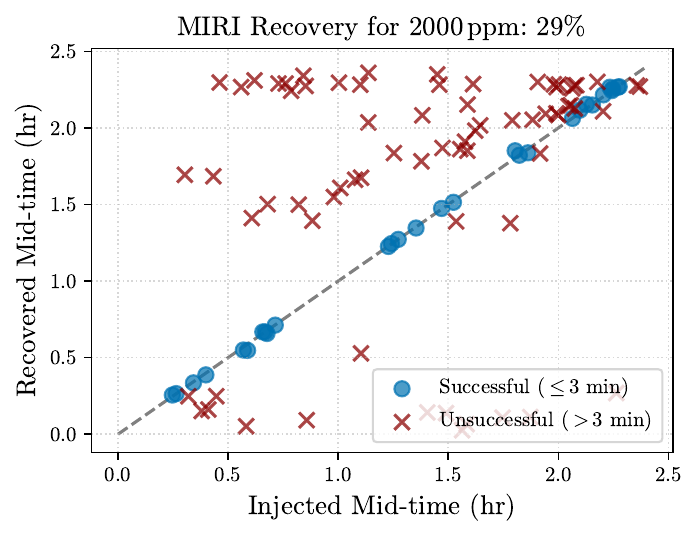} 
        \caption{}
        \label{fig:recall_m2000ppm}
    \end{subfigure}
        \hfill
    \begin{subfigure}[b]{0.329\linewidth}
        \centering
        \includegraphics[width=\linewidth]{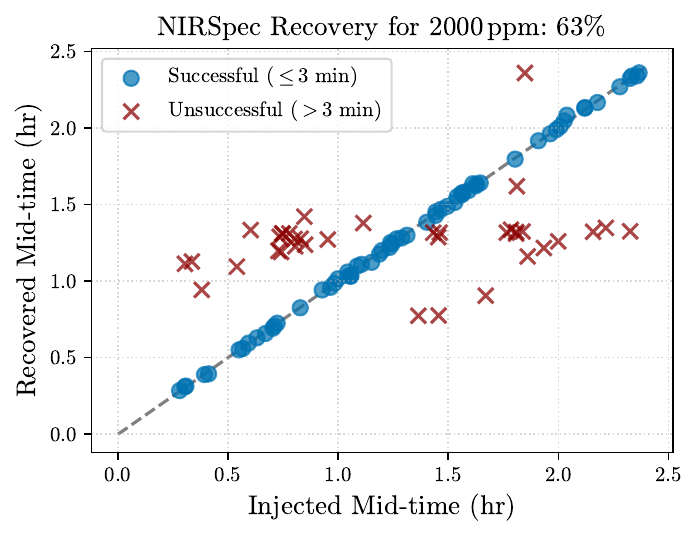}
        \caption{}
        \label{fig:recall_n2000ppm}
    \end{subfigure}
      \begin{subfigure}[b]{0.329\linewidth}
        \centering
        \includegraphics[width=\linewidth]{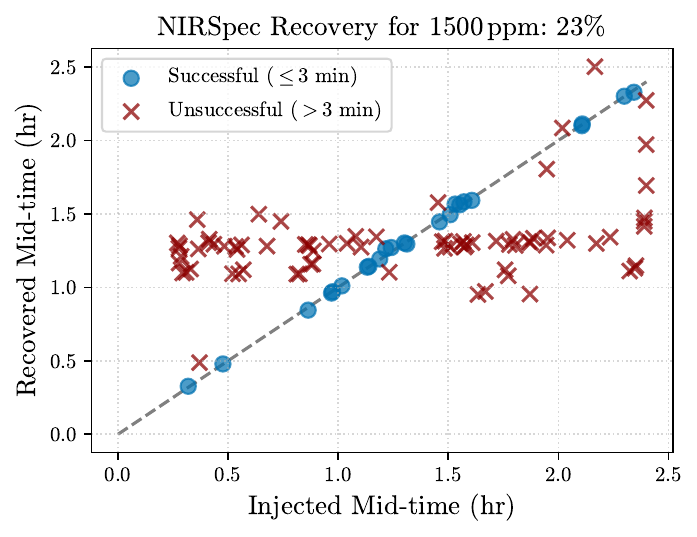}
        \caption{}
        \label{fig:recall_n1500ppm}
    \end{subfigure}
    \caption{Transit injection recovery results for injected transits with depths of 2000\,ppm ($0.61\,\text{R}_\oplus$) for MIRI (panel a) and NIRSpec (panel b) light curves and depths of depths of 3000\,ppm for NIRSpec light curves (panel c). Transit recovery is determined by the recovered mid-time being within $\pm3$\,min of the injected mid-time.}
    \label{fig:recall_comparison}
\end{figure*}

\section{Results}
\label{sec:results}

\subsection{Injection-Recovery Results}
\label{subsec:results_injrec}

We carry out injection-recovery tests in the MIRI and NIRSpec data within the parameter space as described in Sections \ref{subsubsection:inj} and \ref{subsubsection:recovery}. 

Figure \ref{fig:recall_comparison} displays an example of the algorithm's recovery fraction for 100 injections with depths of 2000\,ppm in the MIRI and NIRSpec light curves and 1500\,ppm for NIRSpec. 
A transit recovery is considered successful (blue dot) if the recovered mid-time is within 3 minutes of the injected mid-time. Transits that are not recovered (red $x$) may have been too small (consistent with noise) or lost in the variability features. 

The failed recoveries exhibit different patterns for the MIRI and NIRSpec injections.
For NIRSpec, at small transit depths, the recovery algorithm begins to recover a deep variability feature at approximately 1 hour instead of the injected transit (hence the horizontal line of red $x$'s near recovered mid-time of 1 hour, as shown in Figure~\ref{fig:recall_n2000ppm}). 
This does not persist in our MIRI search as the variability features are different between groups and lower in amplitude (as seen in Figure \ref{fig:bothlcs}); rather, in cases where the algorithm fails to recover transits in the MIRI light curves, the recovered times appear more random (i.e., they do not converge to one specific time/feature, Figure \ref{fig:recall_m2000ppm}).
These results persist in a more direct comparison between MIRI and NIRSpec, where we compare the injection recovery results with roughly the same recovery fraction as shown in Figure \ref{fig:recall_m2000ppm} and Figure \ref{fig:recall_n1500ppm}).

Although the NIRSpec light curves have higher photometric precision than MIRI, their recovery performance will not necessarily be better. In this case, rather than photon noise, the dominant limitation is the high-amplitude variability present in the NIRSpec light curves, whereas this effect is not present to the same degree in the MIRI light curves. Thus, the optimal targets to search for small moons may ideally have low amplitude variability. 

\begin{figure}
    \centering
    \includegraphics[width=0.95\linewidth]{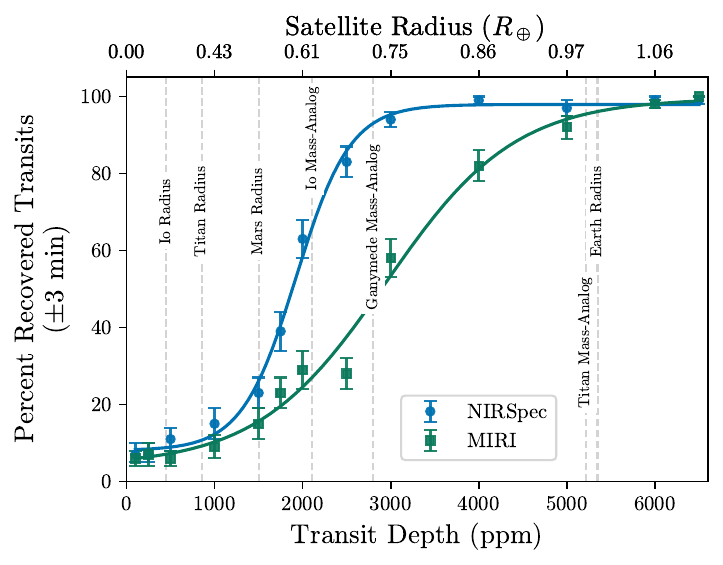}
    \caption{Results of injection-recovery as a function of transit depth in \simp{} light curves. Transit recovery is determined by the recovered mid-time being $\pm3\,$min of the injected mid-time. Green squares show recovery fractions for transits injected in the NIRSpec light curves and blue dots for MIRI. Both are fitted with a logistic function for easier interpretation.%Both contain transits within the physically allowed inclination range 68-90$^\circ$. 
    }
    \label{fig:inj_rec_results}
\end{figure}

The results of the full MIRI and NIRSpec transit injection-recovery tests are displayed in Figure \ref{fig:inj_rec_results} where recovery percentage is shown as a function of injected transit depth. To estimate the uncertainties, we used a non-parametric bootstrap approach.
For each set of 100 injection recoveries at a given transit depth, we generated 10,000 bootstrap samples by resampling the data with replacement and computed the mean recovery rate for each resample. The upper and lower limits were derived from the resulting distribution and correspond to the 16th and 84th percentiles (a 68\% confidence interval). 

For ease of interpretation, the recovery results are fit with logistic function with a floor.
From these fits, we find we are sensitive to (meaning a 50\% recovery threshold) 0.74\,$\text{R}_\oplus$ and 0.60\,$\text{R}_\oplus$ sized transits for MIRI and NIRSpec accordingly. Transits larger than an Earth-radius are almost always recovered, and using NIRSpec injection-recovery results, we are sensitive to 66\% of Io-mass analogs and 93\% of Ganymede-mass analogs. 

Our detection algorithm performs equally (within 1-2$\sigma$) in tests of 100 injections at various depths for MIRI and NIRSpec at the highest and lowest transit depths. However, NIRSpec has up to double the recovery rates than MIRI at intermediate sized transit depths. A longer discussion of the possible causes of this preference and our preference for using NIRSpec in transit searches, even given the high amplitude variability, is found in Section \ref{subsec:comparing_JWST}.

\subsection{Total Transit Detection Probabilities}
\label{subsec:results_probs}

The total probability of observing a transiting exo-Io, $P_{tot}$, using this detection method can be calculated by assessing the geometry of the system, timing of the observation, and the detection algorithm's sensitivity:
\begin{equation}
\label{eqn:Ptot}
    P_{tot}=P_{geo}*P_{obs}*P_{det},
\end{equation}
where $P_{geo}$ is geometric transit probability from Equation \ref{eqn:Pgeo}.
The observational transit probability, $P_{obs}$, is determined by the length of time an object is observed and its orbital period where $P_{obs}= \text{min}(1, \frac{t_{obs duration}}{P})$. The detection probability, $P_{det}$, is the probability the algorithm successfully recovers a transit that is present in the light curve. For \simp, $P_{det}$ is estimated to be the injection-recovery recovery rates at each transit depth shown in Figure \ref{fig:inj_rec_results}.

\begin{figure}
    \centering
    \includegraphics[width=0.9\linewidth]{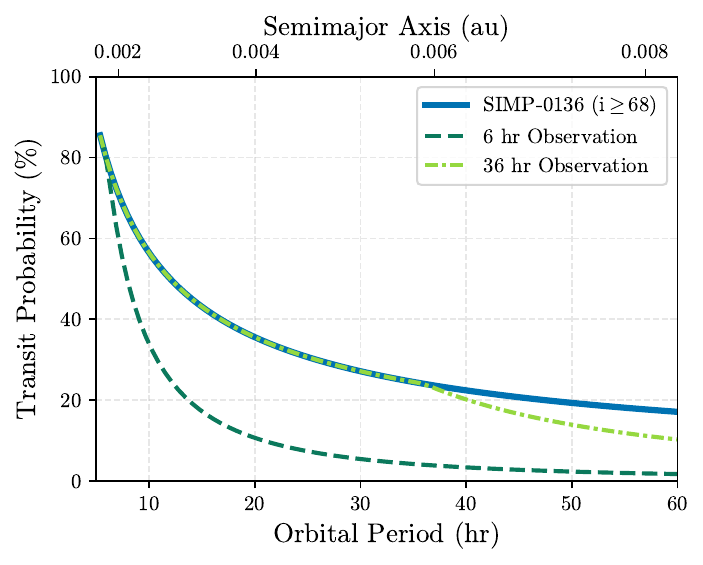}
    \caption{Geometric transit probability ($P_{geo}$, blue line) and the probability of observable transit, ($P_{geo}*P_{obs}$, green dashed lines) for \simp{}, assuming a viewing inclination of $i$ = $68-90^\circ$. }
    \label{fig:geo_obs_prob}
\end{figure}

Before we compute $P_{tot}$, we first compute the product $P_{geo}*P_{obs}$, which we call the probability of observable transit.
We choose a conservative inclination estimate of $i_l=68^{\circ}$ (the lower limit measured by \citealt{vos_2017_viewinggeoBDs}) to calculate the geometric transit probability. 
Figure~\ref{fig:geo_obs_prob} shows the probability of observable transit, $P_{geo}*P_{obs}$, computed for both a 6 hour observation (dark green dashed line; matching the duration of the observations in this study) and a 36 hour observation (light green dashed line), which represents a practical timescale required to place meaningful constraints on the presence of a transiting satellite in this system.

\begin{figure}
    \centering
\includegraphics[width=0.95\linewidth]{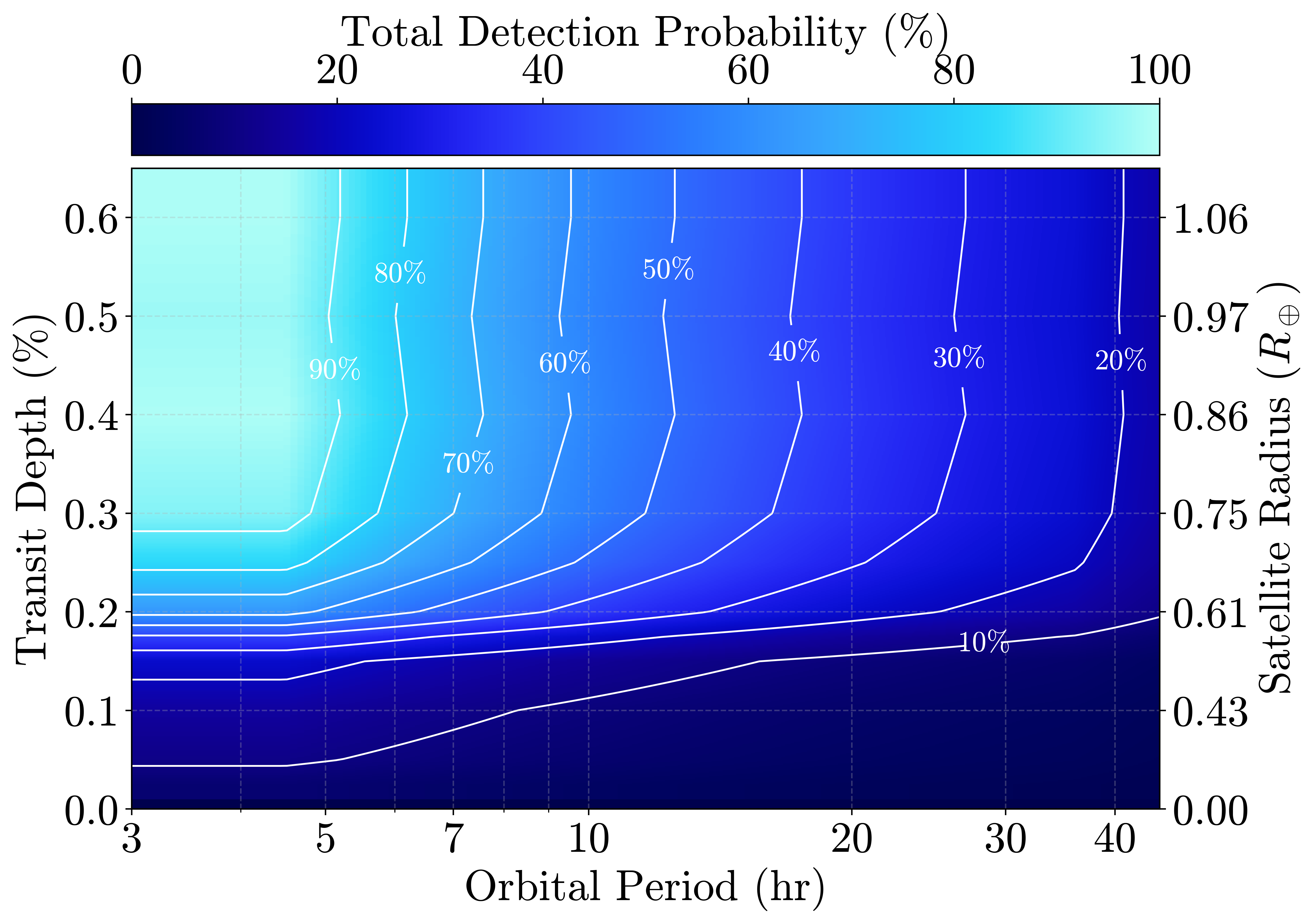}
    \caption{Total probability (Equation \ref{eqn:Ptot}) of observing and detecting an exo-Io transit for a 36 hour observation and system viewing inclination between $i=68-90\deg$, assuming spin-orbit alignment, utilizing  NIRSpec transit recovery rates.}
    \label{fig:p_tot}
\end{figure}

Now we can quantify the full range of exo-Io parameters accessible to our search; we apply the results of our \simp{} analysis within Equation \ref{eqn:Ptot}. Figure \ref{fig:p_tot} displays the total transit probability, $P_{tot}$, of exosatellites (e.g., the $P_{geo}*P_{obs}$ plotted in Figure~\ref{fig:geo_obs_prob} times $P_{det}$ from Figure~\ref{fig:inj_rec_results}) based on the NIRSpec injection-recovery tests.
The plot is smoothed by interpolating between recovery values and their associated orbital periods. 

We find that, if such a satellite is present in the system, the total probability of observing a transit for a Ganymede mass-analog ($0.72\,\text{R}_\oplus$, $0.32\,\text{M}_\oplus$) on a 1 day orbit is $29\%$. 
Thus, theoretically, if every such system hosted a 1 day orbit Ganymede mass-analog, we would only need four 36 hour observations of systems similar to \simp{} (aurorally active isolated substellar objects) before expecting the detection of an exosatellite. 
Conversely, if we observe $\sim$12 systems and detect no satellites, we can rule out their ubiquity (i.e., presence in every system) around aurorally active isolated substellar objects at the $2\sigma$ level. This would provide the first observational evidence that tidally heated exo-Ios are not the primary (or sole) source of the plasma fueling auroral emission from substellar worlds. From our results, we conclude that a systematic search of only a small sample of aurorally active isolated substellar objects would be sufficient to provide meaningful evidence for or against the hypothesis that Io-analog satellites supply the plasma. Such a program would constitute an observationally driven test capable of constraining this scenario.

\section{Discussion}
\label{sec:discussion}

\subsection{Comparing JWST Observing Modes}
\label{subsec:comparing_JWST}

The \simp{} NIRSpec light curve exhibits an amplitude of $\sim$2\% while the MIRI light curves have amplitudes $<$1\%. 
NIRSpec's higher variability amplitudes introduce additional structure on which the transit search algorithm can become trapped and can more easily obscure shallow transit signals. Despite this, NIRSpec outperforms MIRI at the intermediate transit depths in our injection-recovery testing (Figure \ref{fig:inj_rec_results}). 
The superior performance of NIRSpec relative to MIRI is likely attributable to the higher photometric precision in the NIRSpec light curve data (213--359ppm/3min for NIRSpec and 323--553ppm/3min for MIRI). The higher signal-to-noise ratio of the NIRSpec data was predicted by the JWST ETC prior to the observations. We therefore conclude that the smallest transits are likely most readily detectable in wavelength regimes where the signal-to-noise is highest, even when those data exhibit significantly larger variability amplitudes. 

That said, the comparative performance of MIRI is not substantially worse, and the MIRI data remain highly useful for transit searches with nearly comparable sensitivity. For extremely cold substellar worlds (Y-dwarfs), where the near-infrared flux is further reduced relative to the mid-infrared, MIRI may provide superior transit detection limits. However, for early T-dwarfs (such as \simp{}) and L-dwarfs, near-infrared monitoring is likely to yield the most sensitive constraints on transiting exosatellites.

Although NIRSpec outperformed MIRI in the injection recovery tests, we find that our exo-Io search is nevertheless limited by variability. Specifically, the photometric precision of the NIRSpec light curves (213ppm/3min in the highest-precision binned spectral band) should, in principle, enable a $5\sigma$ detection of a $0.27\,R_{\oplus}$ exosatellite (slightly smaller than Io) with a 60~min transit duration. However, satellites of this size are recovered in fewer than 10\% of our injection-recovery trials. While \simp{} is an ideal target for low photon noise and high SNR observations (bright because of its young age and proximity despite its low mass), its fast rotation and high amplitude variability makes it a suboptimal target in both detectors. The best targets will be slow rotators with predictable, low amplitude variability and long baseline observations \citep{Householder2025}.

\subsection{A Targeted JWST Search for Exo-Ios as Auroral Plasma Sources}
\label{subsec:longerLCs}

We have demonstrated the detectability of exosatellites with orbital periods and sizes similar to Jupiter's Io that could be capable of fueling aurora on substellar worlds. However, while JWST can detect transiting exo-Ios, we would not expect to detect an exosatellite in the \simp{} light curve explored in this study due to the short duration of the observation. Specifically, the probability of detecting a transit is low (though not negligible; e.g., a satellite with a $\sim$15~hr orbital period has a transit probability of $\sim$20\%; see Figure~\ref{fig:geo_obs_prob}). Determining whether exo-Ios are common around aurorally active substellar worlds (and therefore represent a plausible plasma source for auroral emission) will require longer duration light curves with state of the art facilities such as JWST or the upcoming ELT and observations of a larger sample of targets.

Specifically, to test whether exo-Ios are common around aurorally active substellar worlds, we must quantify the probability of detecting an exo-Io in a given system. Figure~\ref{fig:geo_obs_prob} shows that for a satellite on a 1.5~day orbit (similar to Io) around \simp{}, the probability of detection is remarkably high, about $25\%$, given the known inclination constraints of the system. 
Therefore a relatively small sample of 4--12 objects similar to \simp{},\footnote{Here, ``similar" implies that the objects have comparable masses and radii and exhibit auroral emission, which may indicate they are viewed closer to edge-on and not pole-on.} each observed for $\sim36$\,hr, would be sufficient to begin placing meaningful constraints on whether exo-Ios are ubiquitous around aurorally active substellar worlds. Such a program would be feasible as a medium- to large-sized JWST survey, using NIRSpec to monitor the light curves of approximately half a dozen aurorally active substellar worlds. Establishing such constraints would be particularly impactful, as the plasma sources powering aurorae in these objects remain poorly understood \citep{pineda_2017}.

In short, we conclude that
this transit search method explored herein provides an opportunity to observationally test if exo-Ios can supply the plasma that fuels aurora on substellar worlds.

\subsection{Theoretical Work Needed to Constrain Exo-Io Populations}
\label{subsec:theoryNeeded}

The population of satellites expected around brown dwarfs and free-floating planetary-mass objects remains poorly explored. While a small number of formation models provide broad expectations for satellite masses and occurrence rates \citep[e.g.,][]{canup2006,liu2020A&A...638A..88L,inderbitzi2020MNRAS.499.1023I,cilibrasi2021}, 
the literature remains sparse, and few studies have specifically predicted the physical and orbital properties of satellites in the substellar host-mass regime. This limits our ability to place detections or non-detections of transiting exosatellites into a broader demographic context and expectations.

Future theoretical work is especially needed to connect satellite formation and orbital evolution to the plasma-source problem. In particular, detailed models could predict the range of satellite sizes, masses, eccentricities, and orbital separations capable of producing sufficient tidal heating \citep{2013ApJ...769...98P} to sustain Io-like or super-Io volcanic activity and generate the plasma outflow needed to fuel aurorae on their hosts. Such work would connect volcanic outflow rates to the plasma densities required to power the observed auroral radio emission \citep{Kao_2018_0136aurora}. A self-consistent treatment of this parameter space would provide physically motivated priors for future transit searches and clarify which exosatellite systems are capable of fueling aurorae. For example, if only very small orbital separations, large satellite sizes, or specific eccentricity ranges can produce sufficient tidal heating and plasma outflow, then searches that are sensitive to that region of parameter space can place strong constraints. 
In that case, a non-detection would not simply indicate the absence of detectable transits, but could instead provide meaningful limits on whether volcanic exosatellites are a viable source of magnetospheric plasma. While this work demonstrates which transiting exosatellites can be detected with current JWST light curves, a more complete theoretical framework is needed to determine the parameter space where such satellites should exist if they are responsible for fueling aurorae.

\subsection{Ingress-like Step in MIRI Light Curves}
\label{subsec:theblip}

During the injection-recovery process, our algorithms consistently recover an anomaly with transit-like ingress-only features in the MIRI light curve. The ingress-like feature appears 2.63 hours into the MIRI observation ($\sim\,6$\,hrs into the total elapsed observations of Figure \ref{fig:bothlcs}) in both groups. 
Our detection code finds a transit depth of 5542\,ppm, which corresponds to a 1.02\,R$_\oplus$ exosatellite ($\sim$1.06\,M$_\oplus$; \citealt{chen_kipping_2017}). 
We detect only an ingress-like feature, but given its proximity to the end of the light curve, it is entirely plausible that the corresponding egress would occur beyond the observational baseline, were this a genuine transit event. 

However, it is also plausible that this ingress event could be due to the known systematic in MIRI light curves as a sudden decrease in flux similar to that found in the JWST/MIRI spectrophotometric phase curve of GJ~1214\,b \citep{Malsky_2025_JWSTblip}. The step in the GJ~1214\,b observations was also similar in appearance to a transit ingress, but it was wavelength-dependent, with the step affecting longest wavelengths the greatest and ranging between a few ppm to 1000\,ppm. The \simp{} MIRI data is too noisy to conclude if the step is wavelength dependent or not. Due to the presence of this MIRI systematic, we consider only transits with both ingress and egress observed in the MIRI data to be strong transit candidates, although we cannot rule out the possibility that this event was caused by a transit.

\section{Conclusion}
\label{sec:conclusions}

In our Solar System, studies suggest that Jupiter's aurorae are supplied with plasma from volcanic outflows on its innermost moon, Io. Indicators of auroral emission have been detected on extrasolar substellar objects. In this work, we investigate whether transiting exosatellites around these aurorally active substellar worlds can be detected, enabling a direct test of whether Io analogs are present in these systems and could plausibly serve as a source of plasma that fuels their aurorae.

Specifically, we assessed the detectability of transiting exosatellites using a short (6 hour) JWST near- and mid-infrared time series observation of \simp{}, a $12.7\,M_J$ ``super-Jupiter'' known to exhibit auroral emission.
We demonstrate sensitivity to exosatellite transits as shallow as 1921\,ppm, corresponding to $\sim$0.60\,$\mathrm{R}_\oplus$ ($\sim$0.16\,$\mathrm{M}_\oplus$; \citealt{chen_kipping_2017}), with a 50\% recovery rate. Because \simp{} is $\sim$13$\times$ more massive than Jupiter, and satellite masses are expected to scale linearly with host mass \citep{canup2006}, our achieved sensitivity corresponds to satellites with mass ratios comparable to the Galilean moons, yielding sensitivity to 66\% of Io-mass analogs and 93\% of Ganymede-mass analogs.

Although the existing light curves demonstrate that the transit technique is capable of detecting exosatellites analogous to Io in the aurorally active \simp{} system, the duration of the archival data is insufficient to place meaningful constraints on the presence of a transiting satellite. We conclude that JWST light curves spanning $\sim$1.5~days for $\sim$4-12 known aurorally active super-Jupiters would be needed to place meaningful statistical constraints whether Io analogs are commonly present in these systems.

\keywords{}

\begin{acknowledgments}

All JWST data used in this paper can be found in
MAST: \dataset[10.17909/pfnd-md36]{http://dx.doi.org/10.17909/pfnd-md36}.
%\url{http://dx.doi.org/10.17909/pfnd-md36}.
This work is based [in part] on observations made with the NASA/ESA/CSA James Webb Space Telescope. The data were obtained from the Mikulski Archive for Space Telescopes at the Space Telescope Science Institute, which is operated by the Association of Universities for Research in Astronomy, Inc., under NASA contract NAS 5-03127 for JWST. These observations are associated with program \#3548.

This material is based upon work where BK was supported by the National Science Foundation Graduate Research Fellowship under Grant No.~DGE 2241144. MK acknowledges support through the Lowell Observatory Early Career Bridge Fund and the Heising-Simons Foundation.
JMV acknowledges support from a Royal Society - Research Ireland University Research Fellowship (URF/1/221932, RF/ERE/221108). JMV, MAS and AM acknowledge support from the European Union through the Exo-PEA ERC project (grant number 101164652). Views and opinions expressed are however those of the author(s) only and do not necessarily reflect those of the European Union or the European Research Council Executive Agency. Neither the European Union nor the granting authority can be held responsible for them.

\end{acknowledgments}

\facilities{JWST(NIRSpec,MIRI)}

\software{{\tt Astropy} \citep{2013A&A...558A..33A, 2018AJ....156..123A, 2022ApJ...935..167A}, {\tt NumPy} \citep{harris2020array}, {\tt dynesty.py}  \citep{2020MNRAS.493.3132S, sergey_koposov_2024_12537467}, {\tt corner.py} \citep{corner},  {\tt Matplotlib} \citep{Hunter:2007}, {\tt pandas} \citep{mckinney-proc-scipy-2010}, {\tt scikit-learn} \citep{scikit-learn}}

\bibliography{main}
\bibliographystyle{aasjournalv7}

\end{document}